\documentclass[review,3p]{elsarticle}

\usepackage{graphicx}
\usepackage{color}
\usepackage{epsfig}
\usepackage{booktabs}
\usepackage{latexsym}
\usepackage{amsmath}
\usepackage{eufrak}
\usepackage{hyperref}
\usepackage{subfigure}
\usepackage{ulem}
\usepackage{lettrine}
\usepackage{verbatim}
\usepackage{array}
\usepackage{varwidth}
\usepackage{mathptmx} 
\usepackage{setspace}
\usepackage{natbib}
\usepackage{rotating}
\usepackage{multirow}
\usepackage{float}
\usepackage{bm}
\usepackage{MnSymbol} 
\usepackage{soul}
\usepackage{upgreek}

\biboptions{numbers, compress}

\journal{Fluids}

\begin{document}

\begin{frontmatter}

\title{Injection of Deformable Capsules in a Reservoir, a Systematic Analysis}

\author[unibas,cemec]{Alessandro Coclite}
\ead{alessandro.coclite@unibas.it}

\author[unibs]{Alberto Gambaruto\corref{cor}}
\ead{alberto.gambaruto@bristol.ac.uk}

\cortext[cor]{Corresponding author}

\address[unibas]{School of Engineering, Universit\`{a} della Basilicata, Viale dell'Ateneo Lucano 10, 85100 Potenza, Italy; alessandro.coclite@unibas.it}

\address[cemec]{Centro di Eccellenza in Meccanica Computazionale (CEMeC), Politecnico di Bari, Via Re David 200 -- 70125 Bari, Italy}

\address[unibs]{Department of Mechanical Engineering, University of Bristol, BS8 1TH Bristol, UK; alberto.gambaruto@bristol.ac.uk (A.M.G.)}

\begin{abstract}

A computational study of capsule ejection from a narrow channel into a reservoir is undertaken for a combination of varying deformable capsule sizes and channel dimensions. A mass-spring membrane model is coupled to an Immersed Boundary--Lattice Boltzmann model solver. The aim of the present work is the description of the capsules motion, deformation and the response of the fluid due to the complex particles' dynamics. The interactions between the capsules affect the local velocity field significantly and are responsible for the dynamics observed. Capsule membrane  deformability is also seen to affect inter-capsule interaction, and we observe that the train of three particles locally homogenizes the velocity field and the leading capsule travels faster than the other two trailing capsules. On the contrary, variations in size of the reservoir do not seem to be relevant, while the ratio of capsule diameter with respect to channel diameter plays a major role as well as the ratio of capsule diameter to inter-capsule spacing. This flow set-up has not been covered in the literature, and consequently we focus on describing capsule motion, membrane deformation and fluid dynamics, as a preliminary investigation in this field.

\end{abstract}


\end{frontmatter}

\section{Introduction}

Haemodynamcis in large arteries is commonly described by the incompressible Newtonian Navier-Stokes equations, hence modelling whole blood to have a constant density and viscosity. While this is acceptable for larger arteries at high flow rates, it is more appropriate to adopt a thixotropic non-Newtonian shear-thinning rheological model for viscosity when a larger variation of shear rates are apparent as evident is smaller vessels or slower flows. This would then take into account the presence of the erythrocytes and other constituents of whole blood \cite{robertson2008, robertson2009} in a continuum model. However, when smaller vessels of the cardiovascular system are considered the dimension of the conduits and the circulating cells are of similar scales, and it is therefore necessary to discretise and model whole blood as a multi-component medium. At these smaller scales, the properties of the cells (such as material constitutive laws for the membrane), the inter-cellular flow interactions (such as flow wakes) and other biochemical and biological phenomena (such as tethering or remodelling), must be considered and together describe a complex physical interplay.

Experimental works have been at the forefront and driving much of the research and our understanding of haemodynamic micro-circulation for many years \cite{goldsmith1971}, and it is only with the advance of commodity computational resources that new numerical methods have been developed and with them computational simulations of micro-circulation have been possible. Numerical simulations provide a fine spatial and temporal resolution of physical variables, which enable for a quantitative analysis and allow for different mathematical models and hypotheses to be tested. Some fundamental studies of cells or capsules have been undertaken using computational simulations, investigating the importance of cell shape and deformability, concentration and apparent viscosity, transport and migration, providing important insight into micro-circulation dynamics \cite{pozrikidis1995, matsunaga2016, tsubota2010, nix2014, omori2012b, matsunaga2015, fedosov20111, fedosov2014}.

Research in the field of blood micro-circulation has seen a range of applications and interests. For example, specially designed micro-channel geometries have been used to separate or sort suspended cells. Such designs include simple, sudden expansions which promote cell focusing \cite{faivre2006, yaginuma2013, faustino2016}, or alternatively repeated sections of hyperbolic micro-channels \cite{rodrigues2016}, wavy channels \cite{di2007}, guiding grooves \cite{hsu2008}, multi-stage micro-fluidic devices involving bends and siphoning \cite{tanaka2012, omori2015}, though a range of different micro-fluidic device configurations exist \cite{pinho2013, bento2018}. These largely make use of inertial forces of the cells \cite{yoon2009, martel2012}, as well as cell deformability \cite{losserand2019, nix2014, yaginuma2013, omori2012}. Interestingly however, micro-channels may also be designed in a very similar fashion to enhance mixing of the flow \cite{sudarsan2006}, and a review of low cost fabrication devices is presented in \cite{faustino2016}.

While inertial effects have been predominantly used to sort cells in micro-fluidic devices, it is known that cell deformability and shape play important roles in their transport dynamics \cite{nix2014, coclite20183, coclite20172, coclite20181, decuzzi2010, omori2012b, omori2015}. The volumetric concentration of suspended particles in flow is also known to affect the apparent viscosity \cite{matsunaga2015, matsunaga2016, fedosov2014} of the medium, and the resulting inter-cellular flow interactions have been observed to affect transport of the cells through different micro-channel configurations \cite{gambaruto2016, gong2009, bessonov2014, vahidkhah2016, sun2006}. The motion of suspended particles in micro-channels are also known to induce a pattern of wall shear stress variation along the wall \cite{xiong2010, freund2014, gambaruto2016}, which is not only important in mechanotransduction and signalling pathways, but also in cell adhesion mechanics \cite{takeishi2016}. The effect of particle suspensions of different sizes has also been investigated, with relevance to leukocyte radial margination \cite{takeishi2014} and micro- and nano-particles on drug delivery \cite{muller2014, takeishi2017}.

In the present work we investigate ejection of capsules from a narrow channel to a reservoir, comparing different size ratios of channel and capsule diameters. Specifically, the aim of the present work is to detail the dynamics of circular capsules when navigating across a geometric discontinuity. The presence of capsules dragged by the flow locally increases the apparent viscosity of the fluid in the region immediately near and inside the membrane itself. This causes the homogenisation of the velocity field disturbed by the presence of the particles, and observe that a train of three particles will tend to act as a single larger body (because of the inter-particle interaction). Finally, the effect of the local increased viscosity also causes the leading capsule to move faster than the other two trailing capsules. We perform numerical simulations of micro-fluidic particulate flow of deformable capsules in discontinuous geometries, with relevance to capsule injection in applications such as drug delivery. This flow set-up has not been covered in the literature, and consequently we focus on describing capsule motion, membrane deformation and fluid dynamics, as a preliminary investigation in this field.

\section{Computational Method}

Computational methods to model and solve for multi-component micro-circulation has developed immensely in the last decades. Numerical methods which discretise the domain as lumped volumes (or masses) of fluid, typically denoted as \textit{particle methods} have been popular, including dissipative particle dynamics (DPD), smoothed particle hydrodynamics (SPH), moving particle semi-implicit method (MPS), multiparticle collision dynamics (MCP) \cite{gambaruto2015,  alizadehrad2012, tanaka2005, noguchi2007, bakhshian2016, bakhshian2019}. These methods are based on expressing the governing equations in a moving reference frame, which is well suited to flows with deformable bodies and moving boundaries. Here we adopt a mixed approach, in which the fluid is solved on a fixed grid, while the capsule membranes are described in a moving reference frame. The solution to the membrane forces is then interpolated to the fixed grid. In doing so we adopt a immersed boundary method, and employ the lattice Boltzmann method as the fluid solver.

\subsection{Lattice Boltzmann Method}

The evolution of the fluid is defined in terms of a set of $N$ discrete distribution functions,$[f_{i}], (i=0,\dots,N-1)$, which obey the dimensionless Boltzmann equation
\begin{equation}
{f_i(\vec{x}+\vec{e}_i\Delta t, t+\Delta t)-f_i(\vec{x}, t)=-\frac{\Delta t}{\tau}[f_i(\vec{x}, t)-f^{eq}_i(\vec{x}, t)]}\, ,
\label{BGK}
\end{equation}
in which $\vec{x}$ and $t$ are the spatial and time coordinates, respectively; $[\vec{e}_{i}],(i=0,...,N-1)$ is the set of $N$ discrete velocities; $\Delta t$ is the time step; and $\tau$ is the relaxation time given by the unique non-null eigenvalue of the collision term in the BGK-approximation \cite{bgk}. The kinematic viscosity of the flow is strictly related to $\tau$ as $\nu=c_s^2\, (\tau-\frac{1}{2}) \Delta t$ being $c_s=\frac{1}{\sqrt{3}}\frac{\Delta x}{\Delta t}$ the reticular speed of sound. The moments of the distribution functions define the fluid density $\rho=\sum _{i}f_{i}$, velocity $\vec{u}= \sum _{i}f_{i} \vec{e}_{i}/ \rho$, and pressure $p=c_{s}^{2} \rho =c_{s}^{2} \sum _{i} f_{i}$. The local equilibrium density functions $[f_{i}^{eq}] (i=0,...,N-1)$ are expressed by the Maxwell-Boltzmann distribution:
\begin{equation}
{f^{eq}_i(\vec{x},t)=\omega_i\rho \Bigl[ 1+\frac{1}{c_s^2}(\vec{e}_i\cdot \vec{u})+\frac{1}{2c_s^4}(\vec{e}_i\cdot \vec{u})^2-\frac{1}{2c_s^2}(\vec{u} \cdot \vec{u}) \Bigr] }\, .
\label{feq}
\end{equation}
On the two-dimensional square lattice with $N=9$ speeds $(D2Q9)$ \cite{d2q9}, the set of discrete velocities is given by 
\begin{equation}
{\vec{e}_i= \begin{cases}
(0,0)\, , & \quad if \quad i = 0 \\
\Biggl(\cos\Biggl(\frac{(i-1)\pi}{2}\Biggr),\sin\Biggl(\frac{(i-1)\pi}{2}\Biggr)\Biggr)\, , & \quad if \quad i = 1-4 \\
\sqrt{2}\Biggl(\cos\Biggl(\frac{(2i-9)\pi}{4}\Biggr),\sin\Biggl(\frac{(2i-9)\pi}{4}\Biggr)\Biggr)\, , & \quad if \quad i = 5-8 \\
\end{cases}}
\label{GaussHermite}
\end{equation}
with the weight,$\omega_i=1/9$ for $i=1-4$, $\omega_i= 1/36$ for $i=5-8$, and $\omega_0=4/9$. Here, we adopt a discretisation in the velocity space of the equilibrium distribution based on the Hermite polynomial expansion of this distribution \cite{shan06bis}.

\subsection{Immersed Boundary Treatment}

Deforming body models are commonly based on continuum approaches using strain energy functions to compute the membrane response \cite{pozrikidis2001,skalak1973,kruger2012}. However, a particle-based model governed by molecular dynamics has emerged due to its mathematical simplicity while providing consistent predictions \cite{dao2006,fedosov20111,nakamura2013,ye2014}. In this work, a particle-based model is employed by coupling the Immersed-Boundary (IB) technique with BGK-lattice Boltzmann solver. The immersed body is a worm-like chain of $nv$ vertices linked with $nl$ linear elements, whose centroids are usually called \textit{Lagrangian markers}. A forcing term $[\vec{\mathcal{F}}_{i}] (i=0,...,8)$, accounting for the immersed boundary, is included as an additional contribution on the right-hand side of Eq.\eqref{BGK}:
\begin{equation}
{f_i(\vec{x}+\vec{e}_i\Delta t, t+\Delta t)-f_i(\vec{x}, t)=-\frac{\Delta t}{\tau}[f_i(\vec{x}, t)-f^{eq}_i(\vec{x}, t)]+\Delta t \vec{\mathcal{F}}_{i}}\, .
\label{forcedBGK}
\end{equation}
$\vec{\mathcal{F}}_{i}$ is expanded in term of the reticular Mach number, $\frac{\vec{e}_{i}}{c_{s}}$, resulting in:
\begin{equation}
{\vec{\mathcal{F}}_{i}=\Biggl(1-\frac{1}{2\tau}\omega_i \Bigl[\frac{\vec{e}_i-\vec{u}}{c_s^2}+\frac{\vec{e}_i\cdot \vec{u}}{c_s^4}\vec{e}_i \Bigr] \Biggr)  \cdot \vec{f}_{ib}}\, ,
\label{ForcingTerm}
\end{equation}
where $\vec{f}_{ib}$ is a body force term. Due to the presence of the forcing term, the mass density and the momentum density are derived as  $\rho=\sum _{i}f_{i}$ and $\rho\vec{u}= \sum _{i}f_{i} \vec{e}_{i}+\frac{\Delta t}{2}\vec{\mathcal{F}}_{i}$.

Within this parametrisation, the forced Navier--Stokes equations is recovered with a second order accuracy \cite{guo2011,derosis2014,derosis20141,suzuki2015,wang2015}. The external boundaries of the computational domain are treated with the known velocity bounce back conditions by Zou and He \cite{zouhe1997}. The IBM procedure, extensively proposed and validated by Coclite and colleagues \cite{coclite20163,coclite20172,coclite20181,coclite20183,coclite20191}, is here adopted and the moving-least squares reconstruction by Vanella et al.~\cite{vanella2009} is employed to exchange all LBM distribution functions between the Eulerian lattice and the Lagrangian chain. Finally, the body force term in Eq.\eqref{ForcingTerm}, $\vec{f}_{ib}$, is evaluated through the formulation by Favier et al~\cite{pinelli2014}.

\textbf{\textit{Elastic Membrane deformation.}} Elastic membranes are modelled by means of an elastic strain, bending resistance, and total enclosed area conservation potentials. Specifically, the nodal forces corresponding to the elastic energy for nodes 1 and 2 connected by edge \textit{l} reads as
\begin{equation}
{\begin{cases}
\vec{F}_{1}^{s}&=-k_{s}(l-l_{0})\frac{\vec{r}_{1,2}}{l} \\
\vec{F}_{2}^{s}&=-k_{s}(l-l_{0})\frac{\vec{r}_{2,1}}{l} \\
\end{cases}}\, ,
\label{strainFor}
\end{equation}
where $\vec{r}_{i,j}=\vec{r}_{i}-\vec{r}_{j}$ with $r_{i}$ position vector of the node \textit{i}.

The bending resistance related to the $v$-th vertex connecting two adjacent element is
\begin{equation}
{V_{v}^{b}=\frac{1}{2}k_{b}(k_{v}-k_{v,0})^{2}}\, ,
\label{bendPot}
\end{equation}
being $k_{b}$ the bending constant, $k_{v}$ the current local curvature in the $v$-th vertex, $k_{v,0}$ the local curvature in the $v$-th vertex for the stress-free configuration. The curvature is evaluated by measuring the variation of the angle between two adjacent elements ($\theta -\theta _{0}$), with $\theta _{0}$ the angle in the stress free configuration. Given this, the forces on the nodes $v_{left}$, $v$, and $v_{right}$ are obtained as
\begin{equation}
{\begin{cases}
\vec{F}_{v_{left}}^{b}&=k_{b}(\theta -\theta_{0})\frac{l_{left}}{l_{left}+l_{right}}\vec{n}_{v} \\
\vec{F}_{v}^{b}&=-k_{b}(\theta -\theta _{0})\vec{n}_{v} \\
\vec{F}_{v_{right}}^{b}&=k_{b}(\theta -\theta_{0})\frac{l_{right}}{l_{left}+l_{right}}\vec{n}_{v} \\
\end{cases}}\, ,
\label{bendFor}
\end{equation}
where $l_{right}$ and $l_{left}$ are the length of the two adjacent left and right edges, respectively, and $\vec{n}_{v}$ is the outward unity vector centred in \textit{v}. Note that, in this context the relation between the strain response constant $k_{s}$ and $k_{b}$ is $E_{b}=\frac{k_{b}}{k_{s}r^{2}}$, where r is the particle radius.

In order to limit the membrane stretching, an effective pressure force term is considered. Thus, the penalty force is expressed in term of the 
reference pressure $p_{ref}$ and directed along the normal inward unity vector of the $l$-th element $n_{l}^{-}$, as 
\begin{equation}
{\vec{F}_{l}^{a}=-k_{a}(1-\frac{A}{A_{0}})p_{ref} \vec{n}_{l}^{-} l_{l}}\, ,
\label{areaFor}
\end{equation}
with $l_{l}$ the length of the selected element, $k_{a}$ the incompressibility coefficient, $A$ the current enclosed area, $A_{0}
$ the enclosed area in the stress-free configuration. The enclosed area is computed using the Green's theorem along the curve, $A=\int{x_{l}dy_{l}-y_{l}dx_{l}}$. Within this formulation $k_{a}=1$ returns a perfectly incompressible membrane. Note that, $\vec{F}_{l}^{a}$ is evenly distributed to the two vertices connecting the \textit{l}-th element ($v_{left}$ and $v_{right}$) as $\vec{F}_{l}^{a}=0.5\vec{F}_{v_{left}}^{a}+0.5\vec{F}_{v_{right}}^{a}$.

\textbf{\textit{Particle-Particle Interaction}} Two-body interactions are modelled with a repulsive potential centred in each vertex. The purely repulsive force is such that the minimum allowed distance between two vertices coming from two different particles is $\Delta x$. The impulse acting on vertex $1$, at a distance $d_{1,2}$ from the vertex $2$ of an adjacent particle, is directed in the inward normal direction identified by $\vec{n}_{1}^{-}$ and is given by:
\begin{equation}
{\vec{F}_{1}^{pp}=\frac{10^{-4}}{8\sqrt{2}}\sqrt{\frac{\Delta x}{d_{1,2}^{5}}}\vec{n}_{1}^{-}}\, .
\label{repFor}
\end{equation}

\textbf{\textit{Hydrodynamics Stresses}} Pressure and viscous stresses exerted by the $l$-th linear element are:
\begin{equation}
{\vec{F}_{l}^{p}(t)=(-p_{l} \vec{n}_{l})l_{l}}\, , \\
\label{hydroForP}
\end{equation}
\begin{equation}
{\vec{F}_{l}^{\tau }(t)=(\bar{\tau }_{l}\cdot \vec{n}_{l})l_{l}}\, ,
\label{hydroForNu}
\end{equation}
where $\bar{\tau }_{l}$ and $p_{l}$ are the viscous stress tensor and the pressure evaluated in the centroid of the element, respectively; 
$\vec{n}_{l}$is the outward normal unit vector while $l_{l}$ is its length. The pressure and velocity derivatives in Eq.s \eqref{hydroForP} and \eqref{hydroForNu} are computed using a probe in the normal positive direction of each element, being the probe length $1.2\Delta x$~\cite{vanella2009,MDdTJCP2016}

\subsection{\textsc{Fluid-Structure interaction}}
Particles dynamics is determined by \textit{dynamics IB} technique described in \cite{coclite20191}, using the solution of the Newton equation for each Lagrangian vertex, accounting for both internal, Eq.s \eqref{strainFor}, \eqref{bendFor}, \eqref{areaFor}, and \eqref{repFor}, and external stresses, Eq.s \eqref{hydroForP} and \eqref{hydroForNu}. Then, no-slip boundary conditions are imposed using a weak coupling approach \cite{coclite20163}. The total force $\vec{F}_{v}^{tot}(t)$ acting on the $v$-th element of the immersed body is evaluated in time and the position of the vertices is updated at each Newtonian dynamics time step considering the membrane mass uniformly 
distributed over the $nv$ vertices,
\begin{equation}
{m_{v}\dot{\vec{u}}_{v}=\vec{F}_{v}^{tot}(t)=\vec{F}_{v}^{s}(t)+\vec{F}_{v}^{b}(t)+\vec{F}_{v}^{a}(t)+\vec{F}_{v}^{pp}(t)+\vec{F}_{v}^{p}(t)+\vec{F}_{v}^{\tau}(t)}\, .
\label{newtonSoft}
\end{equation}

The Newton equation of motion is integrated by using the Verlet algorithm. Specifically, a first tentative velocity is considered into the integration process, $\dot{\vec{x}}_{v,0}(t)$, obtained interpolating the fluid velocity from the surrounding lattice nodes
\begin{equation}
{\vec{x}_{v}(t+\Delta t)=\vec{x}_{v}(t)+\dot{\vec{x}}_{v,0}(t)\Delta t+\frac{1}{2}\frac{\vec{F}_{v}^{tot}(t)}{m_{v}}\Delta t^{2}+O(\Delta t^{3})}\, ,
\label{verletPos}
\end{equation}
then, the velocity at the time level $t+\Delta t$ is computed as
\begin{equation}
{\vec{u}_{v}(t+\Delta t)=\frac{\frac{3}{2}\vec{x}_{v}(t+\Delta t)-2\vec{x}_{v}(t)+\frac{1}{2}\vec{x}_{v}(t-\Delta t)}{\Delta t}+O(\Delta t^{2})}\, ,
\label{verletVel}
\end{equation}
It should be noted that the present formulation is unconditionally stable for small deformation of the capsule membrane and for small velocity variations applied, as previously demonstrated by the authors~\cite{coclite20163,coclite20183,coclite20191}.

\subsection{\textsc{Set-up and Boundary Conditions}}

\begin{figure}[t]
\centering

\includegraphics[scale=0.3]{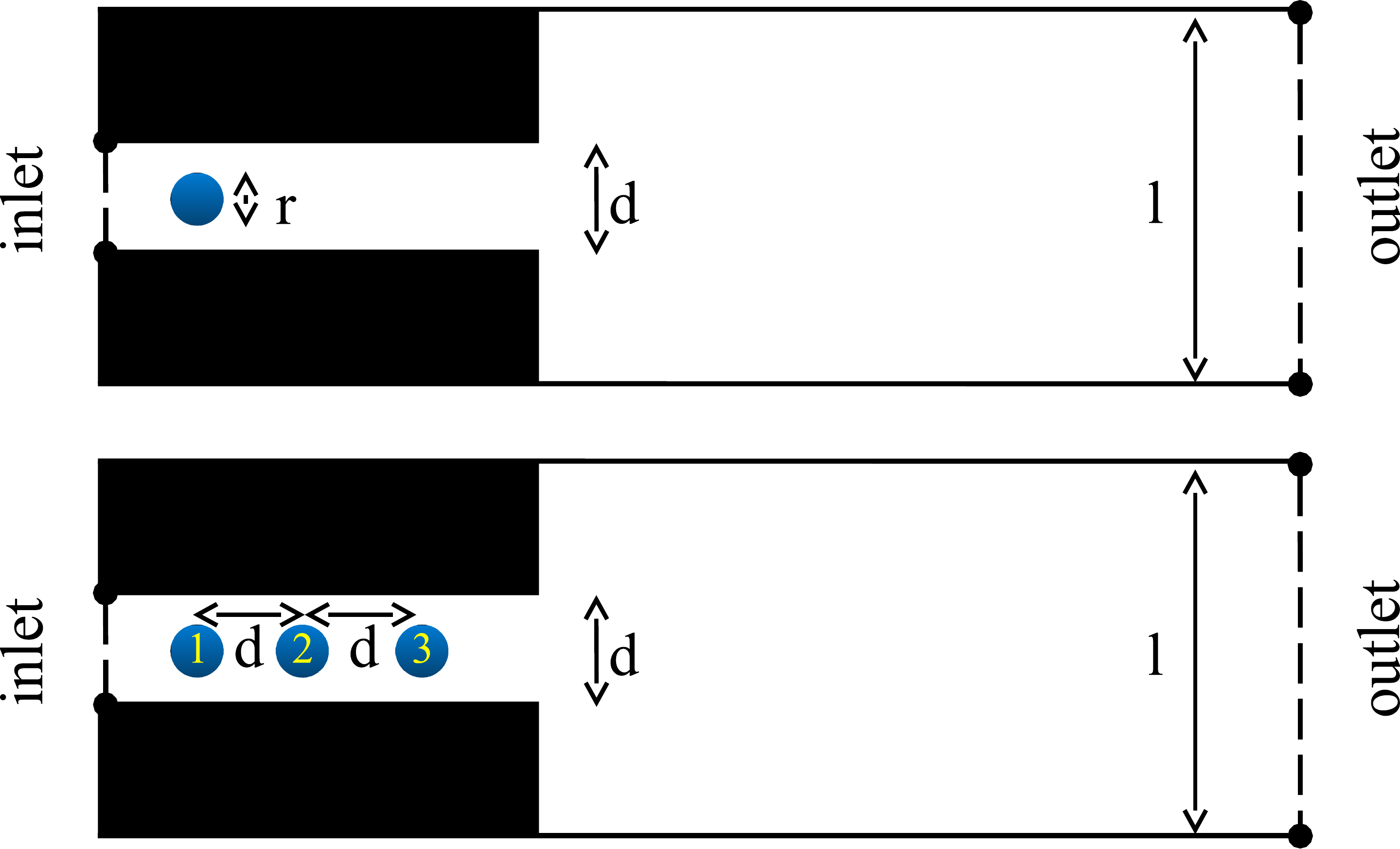}\\
\vspace{3mm}
\begin{tabular}{|c|c|c|c|}
\hline
\textbf{Ca} & \textbf{Re} & \multirow{2}*{\textbf{l/d}} & \multirow{2}*{\textbf{r/d}} \\
($=\nu \, \rho \, u_{max} / k_s$) & ($= u_{max} \, d / \nu$) & & \\
\hline
\multirow{3}*{$10^{-3}$} & \multirow{3}*{$10^{-1}$} & 1.0 & 0.25 \\
& & 2.5 & 0.50 \\
& & 5.0 & 0.75 \\
\hline
\end{tabular}
\caption{{\bf Schematic of the physical problem.} {\bf Top} Sketch of the computational domain with characteristics dimensions and lengths, as well as boundary conditions. {\bf Bottom} Non-dimensional groups used in the computations: the capillary number regulating the mechanical stiffness of the membranes, $Ca$; the Reynolds number regulating the flow velocity, $Re$; the ratio between channel diameter and reservoir height, $l/d$, and the ratio between particle diameter and reservoir height, $r/l$.}
\label{Schematic}
\end{figure} 

The simulations are performed for a two-dimensional domain as shown in Figure~\ref{Schematic} and the fluid is considered to be water. The flow direction is left-to-right, the horizontal axis is denoted by $x-$axis or co-axial direction, and the vertical axis is denoted by $y-$axis or radial direction. The analysis is based on simulations either a single or three in-line spherical capsules, flowing from a channel of small diameter into that of a larger diameter, as shown in Figure~\ref{Schematic}. In subsequent discussion and presentation of results, we refer to the upstream direction as that closer to the inflow, and the downstream direction that closer to the outflow. One should however recognise that the capsule will be travelling faster than the bulk flow in the channel sections, since it is located furthest from the stationary walls. Consequently, in a moving reference frame following a capsule, its wake and disturbance it induces on the flow will in effect be in the upstream direction. 

The computational domain is a rectangular channel by the height $l$ and length $3l$ with no-slip boundary conditions along $y$. The grid resolution is such that $l$ is discretised with $500 \Delta x$. The section $x=0$ is parametrised as a velocity inlet section with a parabolic inlet profile, while the outlet section is located at $x=3l$ (see Figure~\ref{Schematic}) and is set as a convective condition \cite{yang2013}. The Reynolds number is fixed equal to 0.1 and is given by $Re = \frac{u_{max}\, d}{\nu}$; where $u_{max}$ is the maximum velocity for the plane Hagen-Poiseuille profile (parabolic profile) established in $x=0$, $d$ is the narrow channel diameter, and $\nu$ the kinematic viscosity of the fluid ($\nu=1.2\times10^{-6}$~m$^2$/s). Note that, $Re$ represents the ratio of inertial to viscous forces, and can also be interpreted as the ratio of viscous to convective time scales which act on the fluid. Here, the Reynolds number equals 0.1, consequently viscous effects dominate, with viscous forces greater than inertial forces and the viscous time scale is smaller (hence acts faster, stabilising the flow) than the convective time scale. The capsules are initially at the rest, with no pre-stress applied, and of circular section with diameter $r$ and stiffness modulated by the capillary number $Ca = 10^{-3} =\frac{\rho \nu u_{max}}{k_s}$, where $\rho$ is the density of the fluid in which capsules are immersed ($\rho = 1000$~kg/m$^3$) in and $k_s$ is the elastic constant used for the worm-like chain composing the membranes (see Eq~\eqref{strainFor}). $Ca$ represents the ratio of the viscous force to the elastic force consequently representing, within this definition, capsules slightly more rigid than in other studies~\cite{omori2012,nix2014}. Note that, the typical stiffness for a red blood cell is $k_s=1.5\times10^{-4}Nm$ that would lead to Ca$_{rbc}=10^{-2}$ within this definition. To ensure the scheme stability, all the computations are performed with $\tau=1.0$ in the Lattice Boltzmann Method.

The present parametrisation deals with the deformation of a circular membrane in a rectangular two-dimensional channel injected into a reservoir. Given the narrow dimensions and axial symmetry of the problem and set-up, the results obtained are directly transferable to the analogous three-dimensional set-up. For a general case, the deformation and dynamics of initially spherical capsules in a three-dimensional circular capillary flowing into a reservoir will yield different quantitative results. However, the governing physics is unaltered, and while the mechanics in two or three dimensions is different, results from a two-dimensional investigation are transferable to a three-dimensional set-up and one can expect similar qualitative results and trends.

\section{Results and Discussion}

Numerical simulations for flow set-up outlined in Figure~\ref{Schematic} were run for the following three cases: without capsules; with one capsule; with three capsules (numbered left-to-right). The available set-up combinations have resulted in a set of simulations, aimed at sampling the solution space in order to capture the physics of flow of capsules as they are injected into a reservoir.

\begin{figure}[t]
\centering
\includegraphics[scale=0.3]{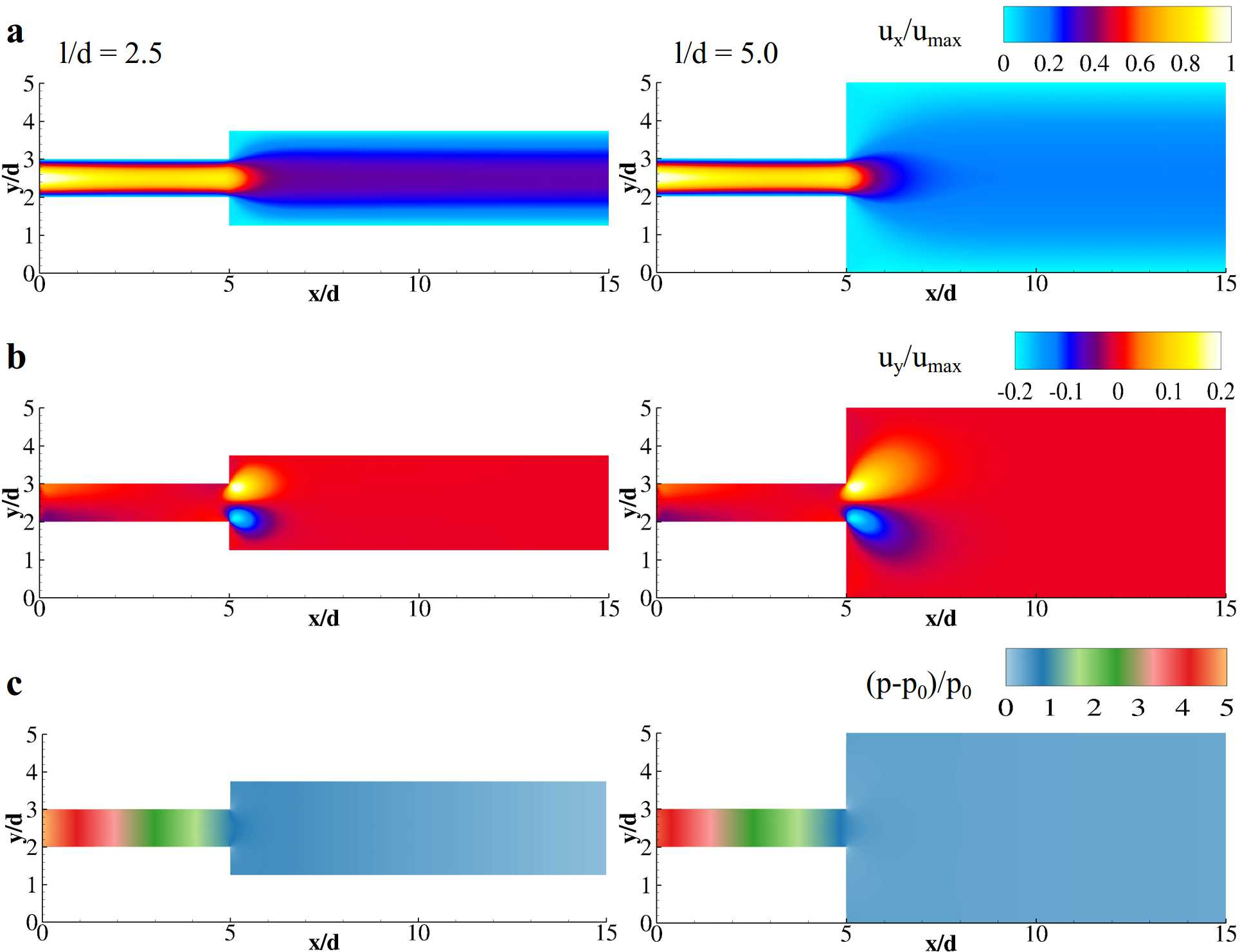}
\caption{{\bf Flow patterns for different width of the reservoir.} {\bf a} Contour of the longitudinal component of the velocity field for $l/d=$ 2.5 (left) and 5.0 (right). {\bf b} Contour of the vertical component of the velocity field for $l/d=$ 2.5 (left) and 5.0 (right). {\bf c} Relative pressure distribution in the computational flow field ($p_0$ is the outlet section pressure). Data for $l/d=$ 1.0 are shown in the Appendix.}
\label{OnlyFluid}
\end{figure} 

The solution for flow without any capsules are shown in Figure~\ref{OnlyFluid} and Figure~\ref{SuppOnlyFluid} (in the Appendix), for purpose of comparison. As expected we see the flow profile develop from the parabolic inflow profile to a flattened paraboloid profile on approaching the reservoir. The flow accelerates in the radial (vertical) direction and decelerates in the co-axial (horizontal) direction as it approaches the end of the smaller channel before discharging into the reservoir. The radial acceleration is effected by a pressure gradient which drives the flow to turn at the geometric discontinuity. The flow separates at the geometric discontinuity, reattaching on the horizontal walls of the larger channel.

\begin{figure}[p]
\centering
\includegraphics[scale=0.25]{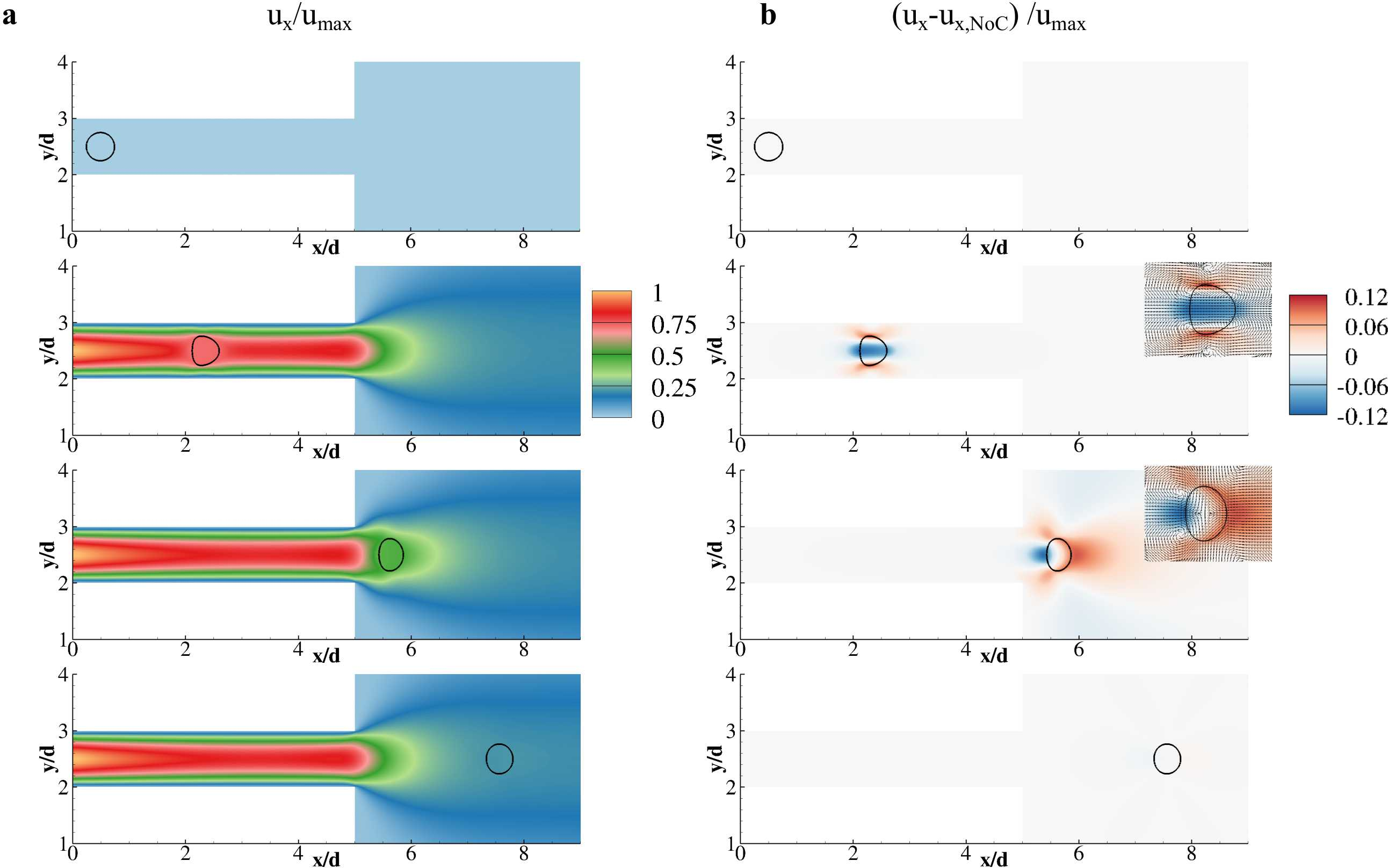}
\caption{{\bf Flow patterns during the transport of a single capsule in the micro-channel.} {\bf a} Contour of the longitudinal velocity field for $l/d=$ 5.0 and $r/d=$ 0.50 taken at four different time steps, namely $t\, u_{max}/d=$ 0, 5, 10, and 20. {\bf b} Contour plot of difference between the longitudinal velocity with ($u_x$) and without ($u_{x,NoC}$) the capsule immersed in at four different time steps, namely $t\, u_{max}/d=$ 0, 5, 10, and 20.} 
\label{PartChipDiff}

\bigskip

\centering
\includegraphics[scale=0.22]{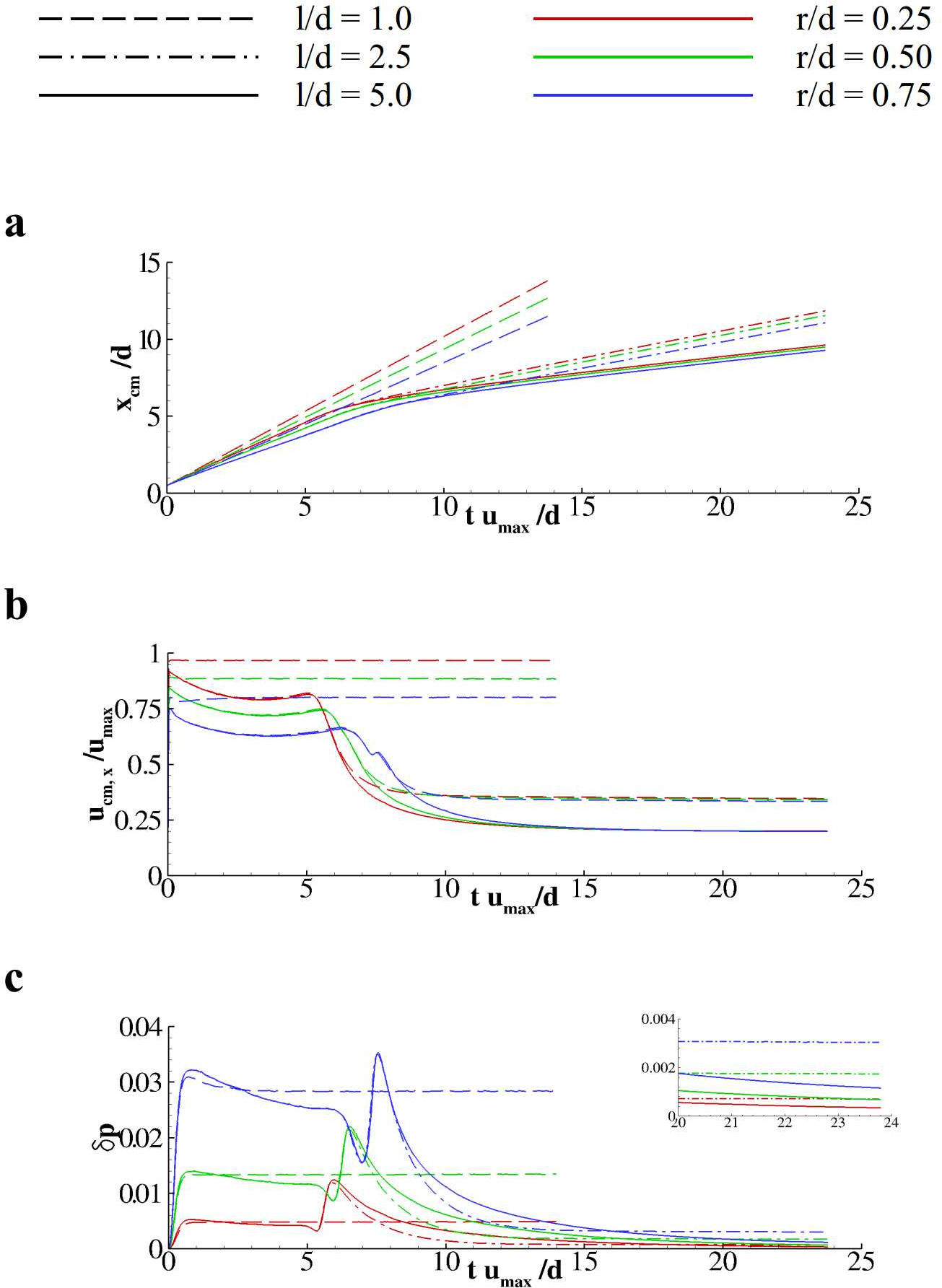}
\caption{{\bf Injection of a capsule in a reservoir.} {\bf a} Distribution of the $x-$coordinate of the capsule centre of mass $x_{cm}$ over time. {\bf b} Distribution of the $x-$velocity of the capsule centre of mass $u_{cm,x}$ over time. {\bf c} Distribution of the capsule stretching computed as the relative difference between the current, $p$, and the initial, $p_0$, perimeter of the capsule ($\delta p=\frac{p-p_0}{p_0}$).} 
\label{SingleCaps}
\end{figure} 

We now turn our attention to the ejection of a single capsule. While the results and discussion are transferable to the other set-up combinations, we present results for the geometry set-up $r/d = 0.50$ and $l/d = 5.0$ in Figure~\ref{PartChipDiff} as these favour presentation and discussion. In Figure~\ref{PartChipDiff} we present the (normalised) co-axial flow component, as well as the difference in flow velocity without capsules and with one capsule (coloured by co-axial velocity component). In the following presentation of results this velocity difference is termed the \textit{relative velocity} or \textit{relative flow field}.

Let us first focus on results for time snapshot $t\, u_{max}/d=5$ in Figure~\ref{PartChipDiff}, when the capsule is located in the narrower channel. We observe that the presence of the capsule tends to create a more uniform velocity profile, by increasing the co-axial velocity near to the walls while decreasing it at the centre of the channel. The vector plot showing the relative velocity also shows that the presence of the flowing capsule results in flow around the capsule, from upstream to downstream. This is because the capsule in effect creates a resistance to the flow, causing it to flow more around it in comparison to the parabolic velocity profile. The relative flow field consequently appears as a vortex pair (or ring in three-dimensions) close to the wall, travelling with the capsule and aligned approximately at its maximum diameter. This vortex ring, identified in the relative flow field, induces a co-axial velocity which drives the capsule through the channel, and this can be seen by the higher velocity at the maximum diameter of the capsule. The wall shear stress is the tangential traction force the flow exerts on the wall due to viscous effects, and we observe that at the capsule's maximum diameter the co-axial velocity near the channel wall is the same as the parabolic flow, and hence the wall shear stress is effectively unaltered. However, slightly upstream and downstream of the capsule's maximum diameter we observe that the co-axial velocity near the wall is greater than for the parabolic profile, indicating a higher wall shear stress. This results in two locations (or rings in three-dimensions) of higher wall shear stress compared to the parabolic flow solution, travelling with the capsule and approximately aligned with the capsule's leading and trailing edges. This is in contract to the results of the wall shear stress `footprint' observed as red blood cells, which show a single higher band of wall shear stress as the deformable cells are in proximity to the walls \cite{xiong2010, freund2014, gambaruto2016}.

We now turn our attention to results for time snapshot $t\, u_{max}/d=10$ in Figure~\ref{PartChipDiff}, when the capsule has been ejected into the reservoir. We observe that for this instance, the mean velocity of the capsule is the same as solution when no capsule is present, while in comparison the flow is moving faster in front of the capsule and slower behind. We also observe that the relative flow field presents a vortex ring at the trailing side of the capsule, travelling with the capsule. This vortex ring is counter-rotating to the vortex ring observed in the narrower channel section, and is set up by the geometric discontinuity. The direction of rotation of this vortex ring induces a velocity which promotes the flow to turn around the geometric discontinuity and results in a smaller flow separation. The induced velocity of this vortex ring also acts to decelerate the capsule co-axial motion as it ejects into the reservoir.

Finally we note that for time snapshot $t\, u_{max}/d= 20$ the flow of a single capsule in a large channel or reservoir has no marked influence on the flow field as compared to the above discussed time snapshots {$t\, u_{max}/d=$ 5 and 10}.

Similar flow fields and relative flow fields were observed in the other set-up combinations. We turn our attention therefore to the motion of the capsule and its change in perimeter length, presented in Figure~\ref{SingleCaps}, and investigate the effects of the geometric variations based on the ratio $l/d$ and $r/d$. Overall the trends appear linear as one varies the geometric ratios $l/d$ and $r/d$, however there are some small deviations which are worth highlighting and discussing.

For comparison purposes in this figure, the case for $l/d=1.0$ (straight channel without discontinuity) is also plotted. We observe that the capsule velocity and perimeter length change are constant (after rapid adjustment from the zero velocity initial conditions of the simulations), however there are variations based on the $r/d$ ratio. At larger values of $r/d$ the capsule forms a larger blockage in the channel, and we observe a more blunt velocity profile is caused by the $r/d=0.75$ capsule compared to the $r/d=0.25$ capsule. The greater perimeter length change is also observed for the $r/d=0.75$ capsule, since there are larger shear rates in proximity with the channel wall aligned to the capsule's largest diameter as it flows. This was observed in Figure~\ref{PartChipDiff} from the relative flow field.

We now focus on the channel geometry set-up with {$l/d=$ 2.5 and 5.0} within Figure~\ref{SingleCaps}. We observe that the traces of normalised co-axial position $x_{cm}/d$ for capsules with {$r/d=$ 0.25 and 0.50} lie closer than for capsule {$r/d = 0.75$}. The lag is seen to appear at the moment when the capsule is injected into the reservoir, with {$t\, u_{max} / d \approx 7.5$}. We observe that the traces of normalised co-axial velocity $u_{cm,x}/u_{max}$ are indistinguishable while the capsule lies within the narrower channel, and parallel based on their $r/d$ value. The capsule velocity then transits to the similar values once the capsule enters the reservoir occurs rapidly, now based on the value of $l/d$, while the ratio $r/d$ has little effect. For $l/d=5.0$ and $r/d=0.75$, we observe that there is a secondary peak in the velocity at {$t\, u_{max} / d \approx 8$}. We observe that the perimeter length change $\delta p$ is relatively constant as the capsule travels along the narrower channel, but decreases and then increases sharply as the capsule is ejected into the reservoir. The decrease in perimeter length is due to the capsule leading edge slowing down as it reaches the reservoir, resulting in a decrease in membrane stress. The subsequent increase in perimeter length occurs as the capsule completes its transition into the reservoir, during which the anterior portion of the capsule is already in the reservoir and has a low velocity, however the posterior portion of the capsule still has a larger velocity, causing the capsule to flatten (stretching radially). Once in the reservoir the capsule membrane relaxes and tends to assume an undeformed shape. The change in perimeter length are larger in the narrower channel due to higher shear rates, more so with increasing $r/d$ ratio, which was also observed from the relative flow field shown in Figure~\ref{PartChipDiff}. For $l/d=5.0$ and $r/d=0.75$, we observe that the sudden decrease and subsequent increase in perimeter lengths were in proportion higher than other cases.

Focusing on the case with $l/d=5.0$ and $r/d=0.75$, we summarise that we observed a different behaviour as the capsule was ejected into the reservoir, compared to the other simulations. This led to a lag in its co-axial position, a second peak in the co-axial velocity and more pronounced change in the perimeter length. The reasons for these phenomena are principally due to the size of the capsule, which owing to the membrane have the effect of locally constraining the flow to be more uniform (hence a homogeneous velocity field). Capsule deformability and the elastic forces are also important, without which we would not obtain the second velocity peak, for example.

\begin{figure}[t]
\centering
\includegraphics[scale=0.25]{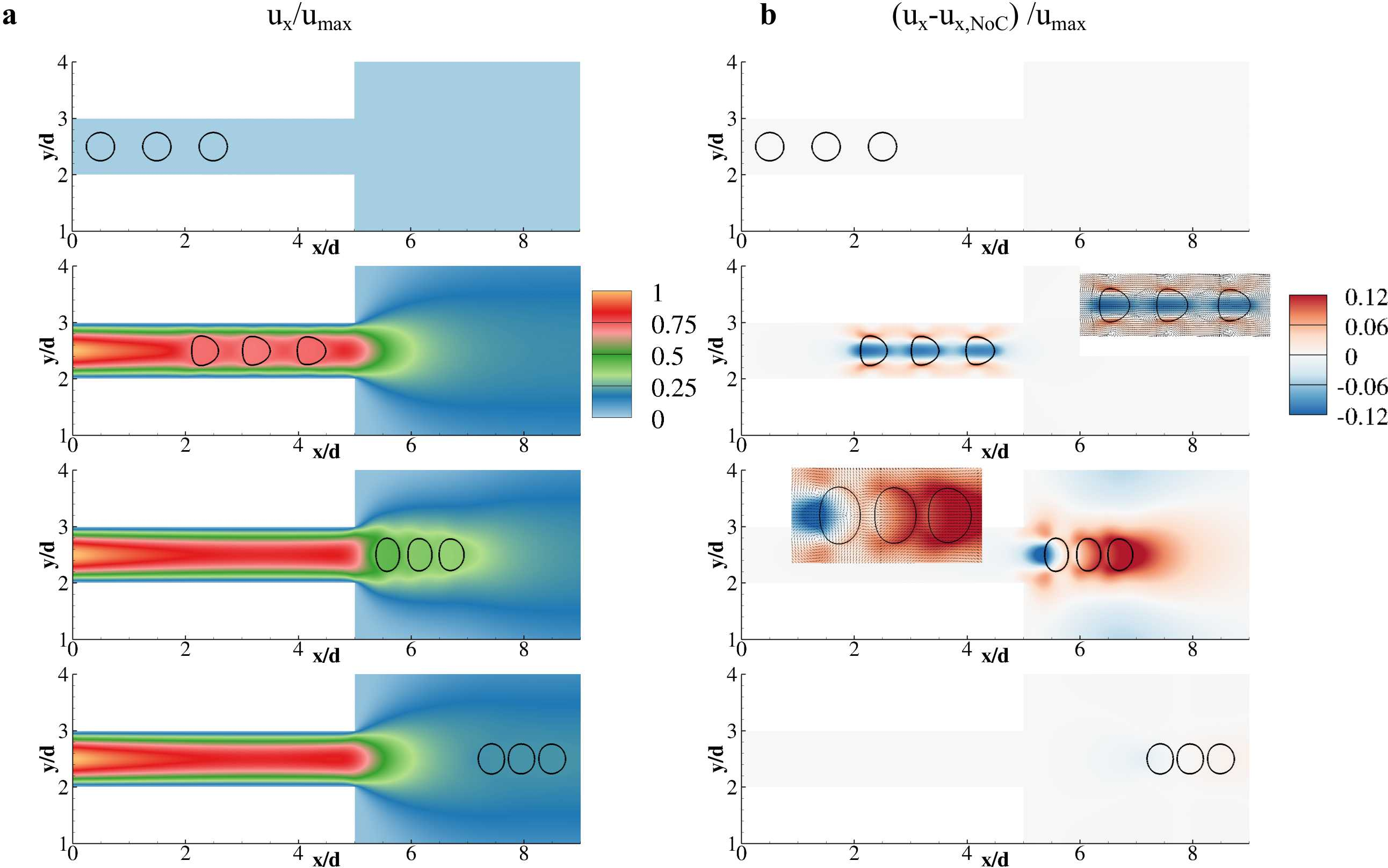}
\caption{{\bf Flow patterns during the transport of three aligned capsules in the micro-channel.} {\bf a} Contour of the longitudinal velocity field for $l/d=$ 5.0 and $r/d=$ 0.50 taken at four different time steps, namely $t u_{max}/d=$ 0, 5, 10, and 20. {\bf b} Contour plot of difference between the longitudinal velocity with ($u_x$) and without ($u_{x,NoC}$) the capsules immersed in at four different time steps, namely $t u_{max}/d=$ 0, 5, 10, and 20.} 
\label{MultiPartChipDiff}
\end{figure} 

We now turn our attention to the results for three capsules flowing in the channel and ejecting into the reservoirs. The flow field and relative velocity for the geometry set-up $r/d = 0.50$ and $l/d = 5.0$ with three capsules is presented in Figure~\ref{MultiPartChipDiff}. In comparison to the flow of the single capsule (see Figure~\ref{PartChipDiff}), we see that within the narrow channel the disturbance in the flow field extends to influence both upstream and downstream capsules, due to their relative proximity. Specifically, the capsules effect a blunter velocity profile with respect to the parabolic profile (obtained without capsules). This phenomenon is often described by a shear-thinning non-Newtonian model for the viscosity. At time $t\, u_{max} / d =10$, when the capsules are entirely in the reservoir, the leading capsules {(capsules 2 and 3)} are travelling faster than is the case without capsules, while the left-most capsule {(capsule 1)} is travelling slower. This is again due to the capsule acting to locally constrain the flow to be more uniform, and in this case extending the apparent jet formed by the flow ejecting from the narrower channel into the reservoir. At this time instance we also note that a vortex ring is formed at the trailing section of capsule~1 only, further highlighting that the flow disturbances between the three capsules result in a local homogenisation of the velocity field, and the capsules therefore tend to act as a single larger object.

\begin{figure}[t]
\centering
\raisebox{0ex}{\begin{turn}{90} \end{turn}}
\includegraphics[scale=0.185]{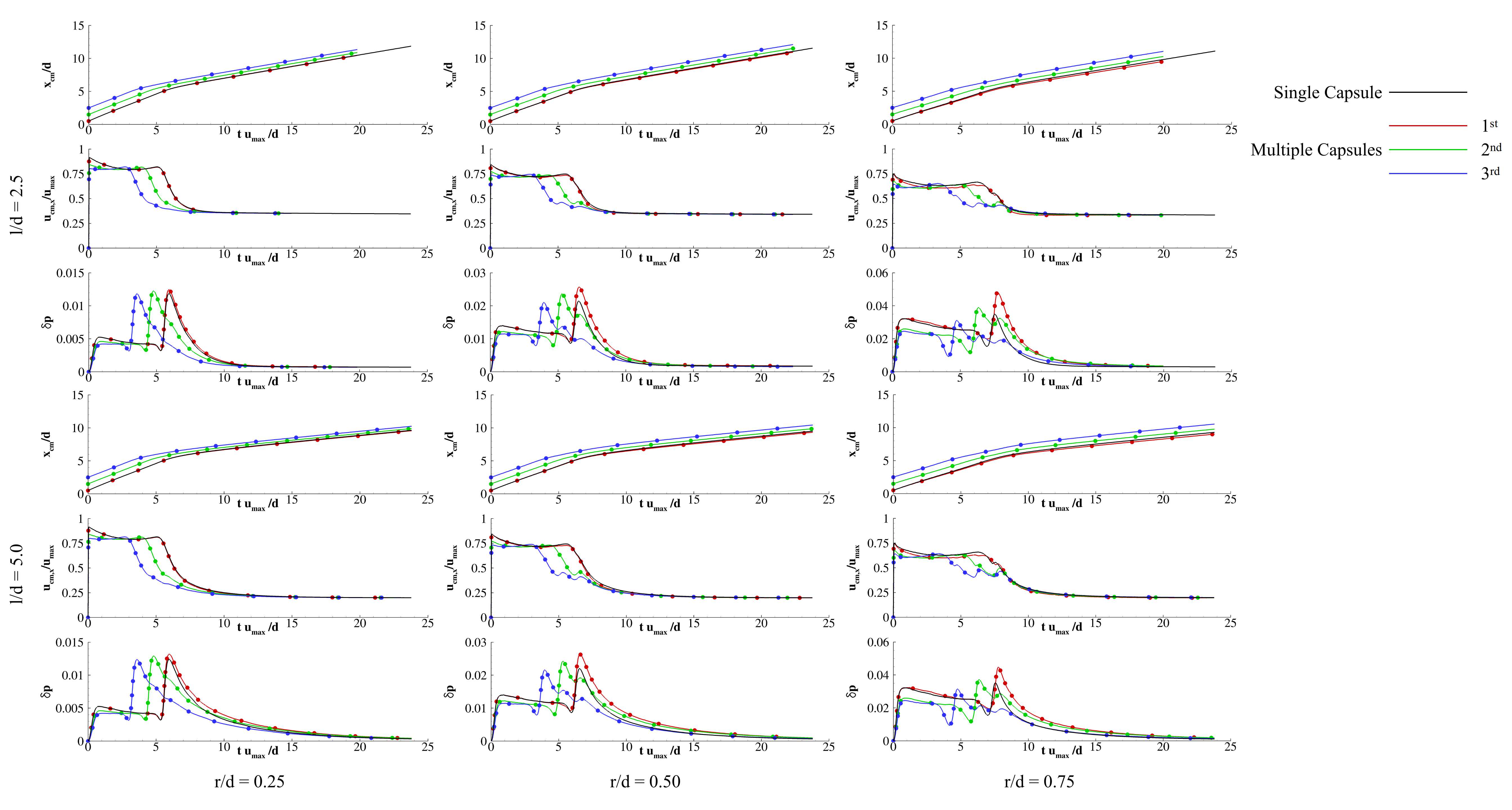} 

\caption{{\bf Transport of three aligned capsules in the micro-channel, for capsule diameter to channel ratio: left column ($r/d=0.25$); middle column ($r/d=0.50$); right column ($r/d=0.75$).} {\bf a} Distribution of the $x-$coordinate of the capsule centre of mass $x_{cm}$ over time for $l/d=2.5$ (top) and $l/d=5.0$ (bottom). {\bf b} Distribution of the $x-$velocity of the capsule centre of mass $u_{cm,x}$ over time for $l/d=2.5$ (top) and $l/d=5.0$ (bottom). {\bf c} Distribution of the capsule stretching computed as the relative difference between the current, $p$, and the initial, $p_0$, perimeter of the capsule ($\delta p=\frac{p-p_0}{p_0}$) for $l/d=2.5$ (top) and $l/d=5.0$ (bottom).} 
\label{MultiCaps_plots}
\end{figure}

In Figure~\ref{MultiCaps_plots} the properties of the single capsule (also shown in Figure~\ref{SingleCaps}) and of the multiple particles are plotted as function of normalised time. We notice that the effect of the relative reservoir diameter, $l/d$, does not play a noticeable influence beyond a given size. Additionally, we observe a greater effect of the multiple capsules when they are larger, and therefore focus our presentation of results for the set-up $l/d = 5.0$ and $r/d=0.75$. The discussion is amenable to the other set-up cases, and differences are presented. In general, for the smaller capsules with {$r/d=$ 0.50 and 0.25} the same phenomena are present as for the larger capsules, but to a lesser extent. Indeed for the smallest capsules $r/d=0.25$, the effect of the multiple cells is almost imperceivable. The reason for this is a reduced inter-capsule interaction, due to the relatively large capsule separation distance and the small capsule size which will not significantly affect the flow field.

In Figure~\ref{MultiCaps_plots}, observing the traces for set-up $l/d = 5.0$ and $r/d=0.75$, we first note that the capsules exhibit oscillating velocities and change in perimeter, more marked for capsule~3 and least for capsule~1. These oscillations are related to the ejection of capsules into the reservoir and we see, for example, that as capsule~1 enters the reservoir it induces an oscillation in both {capsule 2 and 3}. This chain effect is due to the capsules effectively constraining the fluid, and locally homogenising the velocity field. The apparent viscosity is locally higher, and the capsules, due to their close proximity, effectively act as a single larger body. Another phenomenon of particular interest is that capsule~3 tends to move overall faster, and from the $x/d$ trace we see that it is farther from capsules 1 and 2 towards later times. In order to explain this, we turn to Figure~\ref{MultiPartChipDiff}. We observe that the relative velocity is greatest ahead of capsule~3 when the capsules are in the reservoir (at time $t\, u_{max} /d=10$) and the flow is in the narrower channel a short distance ahead of capsule~3 is undisturbed and parabolic. Since capsule~3 is at the leading edge of the capsules, it is not affected by the wake, and will be able to travel faster. This phenomenon of capsule spacing rearrangement was also observed in \cite{di2007}, though in very different geometries. Finally, we note that the perimeter stretching is greatest overall for capsule~1, due to its location in the end of the capsule train, inducing a more marked wake and consequently higher shear rate (velocity gradients) in the fluid and higher stresses in its membrane. In fact we observe that the change in perimeter length is comparable to the set-up of the single capsule, since it has a similar wake flow field.

\clearpage
\newpage
\paperwidth=\pdfpageheight
\paperheight=\pdfpagewidth
\pdfpageheight=\paperheight
\pdfpagewidth=\paperwidth
\newgeometry{layoutwidth=297mm,layoutheight=210 mm, left=2.7cm,right=2.7cm,top=1.8cm,bottom=1.5cm, includehead,includefoot}

\newpage
\restoregeometry
\paperwidth=\pdfpageheight
\paperheight=\pdfpagewidth
\pdfpageheight=\paperheight
\pdfpagewidth=\paperwidth

\section{Conclusions}

In this work we investigate the dynamics of capsule ejection from a narrow channel into a reservoir, across a geometric discontinuity. We observe that inter-capsule interaction (due to the wakes of their motion) and the constraining of the fluid within the membranes are important mechanism which affects the local apparent viscosity since the stress field must be continuous in the domain.

In order to span a meaningful parameter space, a combination of different configurations were investigated. The capsules were varied to have different sizes, namely {$r/d=$ 0.25, 0.50 and 0.75}, where $r$ is the capsule diameter and $d$ is the narrow channel diameter. Three configurations of channel geometries were investigated, namely {$l/d =$ 1.0, 2.5 and 5.0}, where $l$ is the diameter of the reservoir. Additionally, three different configurations: no capsule, a single capsule, and three in-line capsules, were simulated and investigated. The Capillary number and Reynolds numbers were chosen to be $Ca = 10^{-3}$ and $Re=10^{-1}$.

The simulations were investigated by observing the relative flow field, that is the flow field resulting from capsule flow as compared to the no capsule solutions. This has proved to be an effective means of identifying where the flow field has altered, and consequently to identify the fluid mechanics phenomena causing the changes observed. Additionally, the trajectories, velocities and perimeters of the capsules were tracked during the simulations.

Overall we have seen that the reservoir diameter has negligible effect beyond a threshold, and in the resent investigation similar results were obtained for $l/d=2.5$ and $l/d=5.0$. The effect of capsule size was seen to be have a greater effect, with $r/d=0.75$ unsurprisingly resulting in the greatest deviation from a flow field with no capsule, however capsules with size $r/d=0.25$ were also seen to affect the flow field.

Capsule membranes constrain the flow internally, and since the stress field must be continuous across the capsule membrane, the effect is to locally homogenise (i.e. create greater uniformity) the velocity field. This can be seen as a local increase in apparent viscosity. When multiple capsules were investigated, the inter-capsule interaction, caused the capsules to effectively act as a single larger body. This resulted in an increased apparent viscosity spanning the region of the capsules. This effect was clearly observed as the capsules flow in the narrow channel, for which the apparent viscosity resembled that of a shear-thinning non-Newtonian rheological model. An effect of the local increased viscosity is also the cause that the leading capsule tends to move faster than the other two trailing capsules.

The effect of the multiple capsules is to reduce the perimeter change, due to their wakes and inter-capsule interaction which reduces the shear rate (i.e. velocity gradients) of the fluid integrated over the capsule surface. This then leads to a decrease in overall strain for the capsule membrane. The capsule at the trailing edge however is not shielded and its wake promotes a vortex ring in the relative velocity field, and its perimeter change is the same as that of a single capsule flow.

Lastly, we highlight that while the two-dimensional results reported here are representative of the analogous three-dimensional problem, due to the symmetry and regimes (based on capillary and Reynolds numbers) of the set-up, this is generally not the case. Indeed, in complex systems such as a general set-up where deformable capsules are injected into a reservoir, not only is the mechanics of the jet collapse different between two and three dimensions, importantly also the specific stresses involved in the fluid-structure interactions will differ. This noted, two-dimensional simulations can still provide fruitful information on the regulating biophysical mechanisms without the inconvenience of the computationally intense three-dimensional simulations. The extension of the current physical problem to three-dimensional modelling is certainly of interest and will be the object of future investigations.

\section*{References}
\bibliographystyle{unsrtnat}

\begin{thebibliography}{66}
\providecommand{\natexlab}[1]{#1}
\providecommand{\url}[1]{\texttt{#1}}
\expandafter\ifx\csname urlstyle\endcsname\relax
  \providecommand{\doi}[1]{doi: #1}\else
  \providecommand{\doi}{doi: \begingroup \urlstyle{rm}\Url}\fi

\bibitem[Robertson et~al.(2008)Robertson, Sequeira, and
  Kameneva]{robertson2008}
Anne~M Robertson, Ad{\'e}lia Sequeira, and Marina~V Kameneva.
\newblock Hemorheology.
\newblock In \emph{Hemodynamical flows}, pages 63--120. Springer, 2008.

\bibitem[Robertson et~al.(2009)Robertson, Sequeira, and Owens]{robertson2009}
Anne~M Robertson, Ad{\'e}lia Sequeira, and Robert~G Owens.
\newblock Rheological models for blood.
\newblock In \emph{Cardiovascular mathematics}, pages 211--241. Springer, 2009.

\bibitem[Goldsmith(1971)]{goldsmith1971}
HL~Goldsmith.
\newblock Red cell motions and wall interactions in tube flow.
\newblock In \emph{Federation proceedings}, volume~30, pages 1578--1590, 1971.

\bibitem[Pozrikidis(1995)]{pozrikidis1995}
C~Pozrikidis.
\newblock Finite deformation of liquid capsules enclosed by elastic membranes
  in simple shear flow.
\newblock \emph{Journal of Fluid Mechanics}, 297:\penalty0 123--152, 1995.

\bibitem[Matsunaga et~al.(2016)Matsunaga, Imai, Yamaguchi, and
  Ishikawa]{matsunaga2016}
D~Matsunaga, Y~Imai, T~Yamaguchi, and T~Ishikawa.
\newblock Rheology of a dense suspension of spherical capsules under simple
  shear flow.
\newblock \emph{Journal of Fluid Mechanics}, 786:\penalty0 110--127, 2016.

\bibitem[Tsubota and Wada(2010)]{tsubota2010}
Ken-ichi Tsubota and Shigeo Wada.
\newblock Effect of the natural state of an elastic cellular membrane on
  tank-treading and tumbling motions of a single red blood cell.
\newblock \emph{Physical Review E}, 81\penalty0 (1):\penalty0 011910, 2010.

\bibitem[Nix et~al.(2014)Nix, Imai, Matsunaga, Yamaguchi, and
  Ishikawa]{nix2014}
S~Nix, Y~Imai, D~Matsunaga, T~Yamaguchi, and T~Ishikawa.
\newblock Lateral migration of a spherical capsule near a plane wall in stokes
  flow.
\newblock \emph{Physical Review E}, 90\penalty0 (4):\penalty0 043009, 2014.

\bibitem[Omori et~al.(2012{\natexlab{a}})Omori, Imai, Yamaguchi, and
  Ishikawa]{omori2012b}
Toshihiro Omori, Yohsuke Imai, Takami Yamaguchi, and Takuji Ishikawa.
\newblock Reorientation of a nonspherical capsule in creeping shear flow.
\newblock \emph{Physical review letters}, 108\penalty0 (13):\penalty0 138102,
  2012{\natexlab{a}}.

\bibitem[Matsunaga et~al.(2015)Matsunaga, Imai, Yamaguchi, and
  Ishikawa]{matsunaga2015}
D~Matsunaga, Y~Imai, T~Yamaguchi, and T~Ishikawa.
\newblock Deformation of a spherical capsule under oscillating shear flow.
\newblock \emph{Journal of Fluid Mechanics}, 762:\penalty0 288--301, 2015.

\bibitem[Fedosov et~al.(2011)Fedosov, Lei, Caswell, Suresh, and
  Karniadakis]{fedosov20111}
Dmitry~A Fedosov, Huan Lei, Bruce Caswell, Subra Suresh, and George~E
  Karniadakis.
\newblock Multiscale modeling of red blood cell mechanics and blood flow in
  malaria.
\newblock \emph{PLoS computational biology}, 7\penalty0 (12):\penalty0
  e1002270, 2011.

\bibitem[Fedosov et~al.(2014)Fedosov, Noguchi, and Gompper]{fedosov2014}
Dmitry~A Fedosov, Hiroshi Noguchi, and Gerhard Gompper.
\newblock Multiscale modeling of blood flow: from single cells to blood
  rheology.
\newblock \emph{Biomechanics and modeling in mechanobiology}, 13\penalty0
  (2):\penalty0 239--258, 2014.

\bibitem[Faivre et~al.(2006)Faivre, Abkarian, Bickraj, and Stone]{faivre2006}
Magalie Faivre, Manouk Abkarian, Kimberly Bickraj, and Howard~A Stone.
\newblock Geometrical focusing of cells in a microfluidic device: an approach
  to separate blood plasma.
\newblock \emph{Biorheology}, 43\penalty0 (2):\penalty0 147--159, 2006.

\bibitem[Yaginuma et~al.(2013)Yaginuma, Oliveira, Lima, Ishikawa, and
  Yamaguchi]{yaginuma2013}
Tomoko Yaginuma, M{\'o}nica~SN Oliveira, Rui Lima, Takuji Ishikawa, and Takami
  Yamaguchi.
\newblock Human red blood cell behavior under homogeneous extensional flow in a
  hyperbolic-shaped microchannel.
\newblock \emph{Biomicrofluidics}, 7\penalty0 (5):\penalty0 054110, 2013.

\bibitem[Faustino et~al.(2016)Faustino, Catarino, Lima, and
  Minas]{faustino2016}
Vera Faustino, Susana~O Catarino, Rui Lima, and Gra{\c{c}}a Minas.
\newblock Biomedical microfluidic devices by using low-cost fabrication
  techniques: A review.
\newblock \emph{Journal of biomechanics}, 49\penalty0 (11):\penalty0
  2280--2292, 2016.

\bibitem[Rodrigues et~al.(2016)Rodrigues, Lopes, Pinho, Pereira, Garcia,
  Gassmann, Sousa, and Lima]{rodrigues2016}
Raquel~O Rodrigues, Raquel Lopes, Diana Pinho, Ana~I Pereira, Valdemar Garcia,
  Stefan Gassmann, Patr{\'\i}cia~C Sousa, and Rui Lima.
\newblock In vitro blood flow and cell-free layer in hyperbolic microchannels:
  Visualizations and measurements.
\newblock \emph{BioChip Journal}, 10\penalty0 (1):\penalty0 9--15, 2016.

\bibitem[Di~Carlo et~al.(2007)Di~Carlo, Irimia, Tompkins, and Toner]{di2007}
Dino Di~Carlo, Daniel Irimia, Ronald~G Tompkins, and Mehmet Toner.
\newblock Continuous inertial focusing, ordering, and separation of particles
  in microchannels.
\newblock \emph{Proceedings of the National Academy of Sciences}, 104\penalty0
  (48):\penalty0 18892--18897, 2007.

\bibitem[Hsu et~al.(2008)Hsu, Di~Carlo, Chen, Irimia, and Toner]{hsu2008}
Chia-Hsien Hsu, Dino Di~Carlo, Chihchen Chen, Daniel Irimia, and Mehmet Toner.
\newblock Microvortex for focusing, guiding and sorting of particles.
\newblock \emph{Lab on a Chip}, 8\penalty0 (12):\penalty0 2128--2134, 2008.

\bibitem[Tanaka et~al.(2012)Tanaka, Ishikawa, Numayama-Tsuruta, Imai, Ueno,
  Matsuki, and Yamaguchi]{tanaka2012}
Tatsuya Tanaka, Takuji Ishikawa, Keiko Numayama-Tsuruta, Yohsuke Imai, Hironori
  Ueno, Noriaki Matsuki, and Takami Yamaguchi.
\newblock Separation of cancer cells from a red blood cell suspension using
  inertial force.
\newblock \emph{Lab on a Chip}, 12\penalty0 (21):\penalty0 4336--4343, 2012.

\bibitem[Omori et~al.(2015)Omori, Imai, Kikuchi, Ishikawa, and
  Yamaguchi]{omori2015}
Toshihiro Omori, Yohsuke Imai, Kenji Kikuchi, Takuji Ishikawa, and Takami
  Yamaguchi.
\newblock Hemodynamics in the microcirculation and in microfluidics.
\newblock \emph{Annals of biomedical engineering}, 43\penalty0 (1):\penalty0
  238--257, 2015.

\bibitem[Pinho et~al.(2013)Pinho, Yaginuma, and Lima]{pinho2013}
Diana Pinho, Tomoko Yaginuma, and Rui Lima.
\newblock A microfluidic device for partial cell separation and deformability
  assessment.
\newblock \emph{BioChip Journal}, 7\penalty0 (4):\penalty0 367--374, 2013.

\bibitem[Bento et~al.(2018)Bento, Rodrigues, Faustino, Pinho, Fernandes,
  Pereira, Garcia, Miranda, and Lima]{bento2018}
David Bento, Raquel Rodrigues, Vera Faustino, Diana Pinho, Carla Fernandes, Ana
  Pereira, Valdemar Garcia, Jo{\~a}o Miranda, and Rui Lima.
\newblock Deformation of red blood cells, air bubbles, and droplets in
  microfluidic devices: Flow visualizations and measurements.
\newblock \emph{Micromachines}, 9\penalty0 (4):\penalty0 151, 2018.

\bibitem[Yoon et~al.(2009)Yoon, Ha, Bahk, Arakawa, Shoji, and Go]{yoon2009}
Dong~Hyun Yoon, Jin~Bong Ha, Yoen~Kyung Bahk, Takahiro Arakawa, Shuichi Shoji,
  and Jeung~Sang Go.
\newblock Size-selective separation of micro beads by utilizing secondary flow
  in a curved rectangular microchannel.
\newblock \emph{Lab on a Chip}, 9\penalty0 (1):\penalty0 87--90, 2009.

\bibitem[Martel and Toner(2012)]{martel2012}
Joseph~M Martel and Mehmet Toner.
\newblock Inertial focusing dynamics in spiral microchannels.
\newblock \emph{physics of fluids}, 24\penalty0 (3):\penalty0 032001, 2012.

\bibitem[Losserand et~al.(2019)Losserand, Coupier, and
  Podgorski]{losserand2019}
Sylvain Losserand, Gwennou Coupier, and Thomas Podgorski.
\newblock Migration velocity of red blood cells in microchannels.
\newblock \emph{Microvascular research}, 2019.

\bibitem[Omori et~al.(2012{\natexlab{b}})Omori, Ishikawa, Barth\`es-Biesel,
  Salsac, Imai, and Yamaguchi]{omori2012}
T.~Omori, T.~Ishikawa, D.~Barth\`es-Biesel, A.-V. Salsac, Y.~Imai, and
  T.~Yamaguchi.
\newblock Tension of red blood cell membrane in simple shear flow.
\newblock \emph{Phys. Rev. E}, 86:\penalty0 056321, Nov 2012{\natexlab{b}}.
\newblock \doi{10.1103/PhysRevE.86.056321}.
\newblock URL \url{https://link.aps.org/doi/10.1103/PhysRevE.86.056321}.

\bibitem[Sudarsan and Ugaz(2006)]{sudarsan2006}
Arjun~P Sudarsan and Victor~M Ugaz.
\newblock Multivortex micromixing.
\newblock \emph{Proceedings of the National Academy of Sciences}, 103\penalty0
  (19):\penalty0 7228--7233, 2006.

\bibitem[Coclite et~al.(2018)Coclite, Pascazio, de~Tullio, and
  Decuzzi]{coclite20183}
A.~Coclite, G.~Pascazio, M.D. de~Tullio, and P.~Decuzzi.
\newblock Predicting the vascular adhesion of deformable drug carriers in
  narrow capillaries traversed by blood cells.
\newblock \emph{Journal of Fluids and Structures}, 82:\penalty0 638 -- 650,
  2018.
\newblock ISSN 0889-9746.
\newblock \doi{https://doi.org/10.1016/j.jfluidstructs.2018.08.001}.
\newblock URL
  \url{http://www.sciencedirect.com/science/article/pii/S0889974618301944}.

\bibitem[Coclite et~al.(2017)Coclite, Mollica, Ranaldo, Pascazio, de~Tullio,
  and Decuzzi]{coclite20172}
A.~Coclite, H.~Mollica, S.~Ranaldo, G.~Pascazio, M.~D. de~Tullio, and
  P.~Decuzzi.
\newblock Predicting different adhesive regimens of circulating particles at
  blood capillary walls.
\newblock \emph{Microfluidics and Nanofluidics}, 21\penalty0 (11):\penalty0
  168, 2017.
\newblock ISSN 1613-4990.
\newblock \doi{10.1007/s10404-017-2003-7}.
\newblock URL \url{https://doi.org/10.1007/s10404-017-2003-7}.

\bibitem[Mollica et~al.(2018)Mollica, Coclite, Miali, Pereira, Paleari,
  Manneschi, DeCensi, and Decuzzi]{coclite20181}
Hilaria Mollica, Alessandro Coclite, Marco~E. Miali, Rui~C. Pereira, Laura
  Paleari, Chiara Manneschi, Andrea DeCensi, and Paolo Decuzzi.
\newblock Deciphering the relative contribution of vascular inflammation and
  blood rheology in metastatic spreading.
\newblock \emph{Biomicrofluidics}, 12\penalty0 (4):\penalty0 042205, 2018.
\newblock \doi{10.1063/1.5022879}.
\newblock URL \url{https://doi.org/10.1063/1.5022879}.

\bibitem[Decuzzi et~al.(2010)Decuzzi, Godin, Tanaka, Lee, Chiappini, Liu, and
  Ferrari]{decuzzi2010}
P~Decuzzi, B~Godin, T~Tanaka, S-Y Lee, C~Chiappini, X~Liu, and M~Ferrari.
\newblock Size and shape effects in the biodistribution of intravascularly
  injected particles.
\newblock \emph{Journal of Controlled Release}, 141\penalty0 (3):\penalty0
  320--327, 2010.

\bibitem[Gambaruto(2016)]{gambaruto2016}
Alberto~M Gambaruto.
\newblock Flow structures and red blood cell dynamics in arteriole of dilated
  or constricted cross section.
\newblock \emph{Journal of biomechanics}, 49\penalty0 (11):\penalty0
  2229--2240, 2016.

\bibitem[Gong et~al.(2009)Gong, Sugiyama, Takagi, and Matsumoto]{gong2009}
Xiaobo Gong, Kazuyasu Sugiyama, Shu Takagi, and Yoichiro Matsumoto.
\newblock The deformation behavior of multiple red blood cells in a capillary
  vessel.
\newblock \emph{Journal of biomechanical engineering}, 131\penalty0
  (7):\penalty0 074504, 2009.

\bibitem[Bessonov et~al.(2014)Bessonov, Babushkina, Golovashchenko,
  Tosenberger, Ataullakhanov, Panteleev, Tokarev, and Volpert]{bessonov2014}
N~Bessonov, Evgenia Babushkina, SF~Golovashchenko, Alen Tosenberger,
  F~Ataullakhanov, M~Panteleev, A~Tokarev, and Vitaly Volpert.
\newblock Numerical modelling of cell distribution in blood flow.
\newblock \emph{Mathematical Modelling of Natural Phenomena}, 9\penalty0
  (6):\penalty0 69--84, 2014.

\bibitem[Vahidkhah et~al.(2016)Vahidkhah, Balogh, and Bagchi]{vahidkhah2016}
Koohyar Vahidkhah, Peter Balogh, and Prosenjit Bagchi.
\newblock Flow of red blood cells in stenosed microvessels.
\newblock \emph{Scientific reports}, 6:\penalty0 28194, 2016.

\bibitem[Sun and Munn(2006)]{sun2006}
Chenghai Sun and Lance~L Munn.
\newblock Influence of erythrocyte aggregation on leukocyte margination in
  postcapillary expansions: a lattice boltzmann analysis.
\newblock \emph{Physica A: Statistical Mechanics and its Applications},
  362\penalty0 (1):\penalty0 191--196, 2006.

\bibitem[Xiong and Zhang(2010)]{xiong2010}
Wenjuan Xiong and Junfeng Zhang.
\newblock Shear stress variation induced by red blood cell motion in
  microvessel.
\newblock \emph{Annals of Biomedical engineering}, 38\penalty0 (8):\penalty0
  2649--2659, 2010.

\bibitem[Freund and Vermot(2014)]{freund2014}
Jonathan~B Freund and Julien Vermot.
\newblock The wall-stress footprint of blood cells flowing in microvessels.
\newblock \emph{Biophysical journal}, 106\penalty0 (3):\penalty0 752--762,
  2014.

\bibitem[Takeishi et~al.(2016)Takeishi, Imai, Ishida, Omori, Kamm, and
  Ishikawa]{takeishi2016}
Naoki Takeishi, Yohsuke Imai, Shunichi Ishida, Toshihiro Omori, Roger~D Kamm,
  and Takuji Ishikawa.
\newblock Cell adhesion during bullet motion in capillaries.
\newblock \emph{American Journal of Physiology-Heart and Circulatory
  Physiology}, 311\penalty0 (2):\penalty0 H395--H403, 2016.

\bibitem[Takeishi et~al.(2014)Takeishi, Imai, Nakaaki, Yamaguchi, and
  Ishikawa]{takeishi2014}
Naoki Takeishi, Yohsuke Imai, Keita Nakaaki, Takami Yamaguchi, and Takuji
  Ishikawa.
\newblock Leukocyte margination at arteriole shear rate.
\newblock \emph{Physiological reports}, 2\penalty0 (6), 2014.

\bibitem[Muller et~al.(2014)Muller, Fedosov, and Gompper]{muller2014}
K.~Muller, D.A. Fedosov, and G.~Gompper.
\newblock Margination of micro- and nano-particles in blood flow and its effect
  on drug delivery.
\newblock \emph{Scientific Reports}, 4, 2014.
\newblock \doi{10.1038/srep04871}.

\bibitem[Takeishi and Imai(2017)]{takeishi2017}
Naoki Takeishi and Yohsuke Imai.
\newblock Capture of microparticles by bolus flow of red blood cells in
  capillaries.
\newblock \emph{Scientific reports}, 7\penalty0 (1):\penalty0 5381, 2017.

\bibitem[Gambaruto(2015)]{gambaruto2015}
Alberto~M Gambaruto.
\newblock Computational haemodynamics of small vessels using the moving
  particle semi-implicit (mps) method.
\newblock \emph{Journal of Computational Physics}, 302:\penalty0 68--96, 2015.

\bibitem[Alizadehrad et~al.(2012)Alizadehrad, Imai, Nakaaki, Ishikawa, and
  Yamaguchi]{alizadehrad2012}
Davod Alizadehrad, Yohsuke Imai, Keita Nakaaki, Takuji Ishikawa, and Takami
  Yamaguchi.
\newblock Quantification of red blood cell deformation at high-hematocrit blood
  flow in microvessels.
\newblock \emph{Journal of biomechanics}, 45\penalty0 (15):\penalty0
  2684--2689, 2012.

\bibitem[Tanaka and Takano(2005)]{tanaka2005}
Nobuatsu Tanaka and Tatsuo Takano.
\newblock Microscopic-scale simulation of blood flow using sph method.
\newblock \emph{International Journal of Computational Methods}, 2\penalty0
  (04):\penalty0 555--568, 2005.

\bibitem[Noguchi and Gompper(2007)]{noguchi2007}
Hiroshi Noguchi and Gerhard Gompper.
\newblock Swinging and tumbling of fluid vesicles in shear flow.
\newblock \emph{Physical review letters}, 98\penalty0 (12):\penalty0 128103,
  2007.

\bibitem[Bhatnagar et~al.(1954)Bhatnagar, Gross, and Krook]{bgk}
P.~L. Bhatnagar, E.~P. Gross, and M.~Krook.
\newblock A model for collision processes in gases. i. small amplitude
  processes in charged and neutral one-component systems.
\newblock \emph{Phys. Rev.}, 94:\penalty0 511--525, May 1954.
\newblock \doi{10.1103/PhysRev.94.511}.

\bibitem[Qian et~al.(1992)Qian, d'Humi{\`e}res, and Lallemand]{d2q9}
Yue-Hong Qian, Dominique d'Humi{\`e}res, and Pierre Lallemand.
\newblock Lattice bgk models for navier-stokes equation.
\newblock \emph{EPL (Europhysics Letters)}, 17\penalty0 (6):\penalty0 479,
  1992.

\bibitem[Shan et~al.(2006)Shan, Yuan, and Chen]{shan06bis}
Xiaowen Shan, Xue-Feng Yuan, and Hudong Chen.
\newblock Kinetic theory representation of hydrodynamics: a way beyond the
  navier–stokes equation.
\newblock \emph{Journal of Fluid Mechanics}, 550:\penalty0 413--441, 3 2006.
\newblock ISSN 1469-7645.
\newblock \doi{10.1017/S0022112005008153}.

\bibitem[Pozrikidis(2001)]{pozrikidis2001}
C~Pozrikidis.
\newblock Effect of membrane bending stiffness on the deformation of capsules
  in simple shear flow.
\newblock \emph{Journal of Fluid Mechanics}, 440:\penalty0 269--291, 2001.

\bibitem[Skalak et~al.(1973)Skalak, Tozeren, Zarda, and Chien]{skalak1973}
R~Skalak, A~Tozeren, RP~Zarda, and S~Chien.
\newblock Strain energy function of red blood cell membranes.
\newblock \emph{Biophysical Journal}, 13\penalty0 (3):\penalty0 245--264, 1973.

\bibitem[Kr{\"u}ger(2012)]{kruger2012}
Heinrich Kr{\"u}ger.
\newblock \emph{Computer simulation study of collective phenomena in dense
  suspensions of red blood cells under shear}.
\newblock Springer Science \& Business Media, 2012.

\bibitem[Dao et~al.(2006)Dao, Li, and Suresh]{dao2006}
M~Dao, J~Li, and S~Suresh.
\newblock Molecularly based analysis of deformation of spectrin network and
  human erythrocyte.
\newblock \emph{Materials Science and Engineering: C}, 26\penalty0
  (8):\penalty0 1232--1244, 2006.

\bibitem[Nakamura et~al.(2013)Nakamura, Bessho, and Wada]{nakamura2013}
Masanori Nakamura, Sadao Bessho, and Shigeo Wada.
\newblock Spring-network-based model of a red blood cell for simulating
  mesoscopic blood flow.
\newblock \emph{International journal for numerical methods in biomedical
  engineering}, 29\penalty0 (1):\penalty0 114--128, 2013.

\bibitem[Ye et~al.(2014)Ye, Ng, Tan, Leo, and Kim]{ye2014}
Swe~Soe Ye, Yan~Cheng Ng, Justin Tan, Hwa~Liang Leo, and Sangho Kim.
\newblock Two-dimensional strain-hardening membrane model for large deformation
  behavior of multiple red blood cells in high shear conditions.
\newblock \emph{Theoretical Biology and Medical Modelling}, 11\penalty0
  (1):\penalty0 19, 2014.

\bibitem[Guo et~al.(2011)Guo, Zheng, and Shi]{guo2011}
Zhaoli Guo, Chuguang Zheng, and Baochang Shi.
\newblock Force imbalance in lattice boltzmann equation for two-phase flows.
\newblock \emph{Phys. Rev. E}, 83:\penalty0 036707, Mar 2011.
\newblock \doi{10.1103/PhysRevE.83.036707}.

\bibitem[De Rosis et~al.(2014)De Rosis, Ubertini, and Ubertini]{derosis2014}
Alessandro De Rosis, Stefano Ubertini, and Francesco Ubertini.
\newblock A comparison between the interpolated bounce-back scheme and the
  immersed boundary method to treat solid boundary conditions for laminar flows
  in the lattice boltzmann framework.
\newblock \emph{Journal of Scientific Computing}, 61\penalty0 (3):\penalty0
  477--489, 2014.
\newblock ISSN 1573-7691.
\newblock \doi{10.1007/s10915-014-9834-0}.
\newblock URL \url{http://dx.doi.org/10.1007/s10915-014-9834-0}.

\bibitem[Rosis et~al.(2014)Rosis, Ubertini, and Ubertini]{derosis20141}
Alessandro~De Rosis, Stefano Ubertini, and Francesco Ubertini.
\newblock A partitioned approach for two-dimensional fluid-structure
  interaction problems by a coupled lattice boltzmann-finite element method
  with immersed boundary.
\newblock \emph{Journal of Fluids and Structures}, 45:\penalty0 202 -- 215,
  2014.
\newblock ISSN 0889-9746.
\newblock \doi{http://dx.doi.org/10.1016/j.jfluidstructs.2013.12.009}.

\bibitem[Suzuki et~al.(2015)Suzuki, Minami, and Inamuro]{suzuki2015}
K.~Suzuki, K.~Minami, and T.~Inamuro.
\newblock Lift and thrust generation by a butterfly-like flapping wing-body
  model: Immersed boundary-lattice boltzmann simulations.
\newblock \emph{Journal of Fluid Mechanics}, 767:\penalty0 659--695, 2015.
\newblock \doi{10.1017/jfm.2015.57}.

\bibitem[Wang et~al.(2015)Wang, Shu, Teo, and Wu]{wang2015}
Y.~Wang, C.~Shu, C.J. Teo, and J.~Wu.
\newblock An immersed boundary-lattice boltzmann flux solver and its
  applications to fluid-structure interaction problems.
\newblock \emph{Journal of Fluids and Structures}, 54:\penalty0 440 -- 465,
  2015.
\newblock ISSN 0889-9746.
\newblock \doi{http://dx.doi.org/10.1016/j.jfluidstructs.2014.12.003}.

\bibitem[Zou and He(1997)]{zouhe1997}
Qisu Zou and Xiaoyi He.
\newblock On pressure and velocity boundary conditions for the lattice
  boltzmann bgk model.
\newblock \emph{Physics of Fluids}, 9\penalty0 (6):\penalty0 1591--1598, 1997.
\newblock \doi{http://dx.doi.org/10.1063/1.869307}.

\bibitem[Coclite et~al.(2016)Coclite, de~Tullio, Pascazio, and
  Decuzzi]{coclite20163}
A.~Coclite, M.~D. de~Tullio, G.~Pascazio, and P.~Decuzzi.
\newblock A combined lattice boltzmann and immersed boundary approach for
  predicting the vascular transport of differently shaped particles.
\newblock \emph{Computers {\&} Fluids}, 136:\penalty0 260 -- 271, 2016.
\newblock ISSN 0045-7930.
\newblock \doi{http://dx.doi.org/10.1016/j.compfluid.2016.06.014}.

\bibitem[Coclite et~al.(2019)Coclite, Ranaldo, de~Tullio, Decuzzi, and
  Pascazio]{coclite20191}
A.~Coclite, S.~Ranaldo, M.D. de~Tullio, P.~Decuzzi, and G.~Pascazio.
\newblock Kinematic and dynamic forcing strategies for predicting the transport
  of inertial capsules via a combined lattice boltzmann immersed boundary
  method.
\newblock \emph{Computers {\&} Fluids}, 180:\penalty0 41--53, 2019.
\newblock ISSN 0045-7930.
\newblock \doi{https://doi.org/10.1016/j.compfluid.2018.12.014}.
\newblock URL
  \url{http://www.sciencedirect.com/science/article/pii/S0045793018304304}.

\bibitem[Vanella and Balaras(2009)]{vanella2009}
Marcos Vanella and Elias Balaras.
\newblock A moving-least-squares reconstruction for embedded-boundary
  formulations.
\newblock \emph{Journal of Computational Physics}, 228\penalty0 (18):\penalty0
  6617 -- 6628, 2009.
\newblock ISSN 0021-9991.
\newblock \doi{http://dx.doi.org/10.1016/j.jcp.2009.06.003}.

\bibitem[Favier et~al.(2014)Favier, Revell, and Pinelli]{pinelli2014}
J.~Favier, A.~Revell, and A.~Pinelli.
\newblock A lattice boltzmann-immersed boundary method to simulate the fluid
  interaction with moving and slender flexible objects.
\newblock \emph{Journal of Computational Physics}, 261:\penalty0 145--161,
  2014.
\newblock \doi{10.1016/j.jcp.2013.12.052}.

\bibitem[de~Tullio and Pascazio(2016)]{MDdTJCP2016}
Marco~D de~Tullio and Giuseppe Pascazio.
\newblock A moving-least-squares immersed boundary method for simulating the
  fluid--structure interaction of elastic bodies with arbitrary thickness.
\newblock \emph{Journal of Computational Physics}, 325:\penalty0 201--225,
  2016.

\bibitem[Yang(2013)]{yang2013}
Zhaoxia Yang.
\newblock Lattice boltzmann outflow treatments: Convective conditions and
  others.
\newblock \emph{Computers \& Mathematics with Applications}, 65\penalty0
  (2):\penalty0 160--171, 2013.

\end{thebibliography}

\newpage
\section{Supplementary Figures}

\begin{figure}[h]
\centering
\includegraphics[scale=0.3]{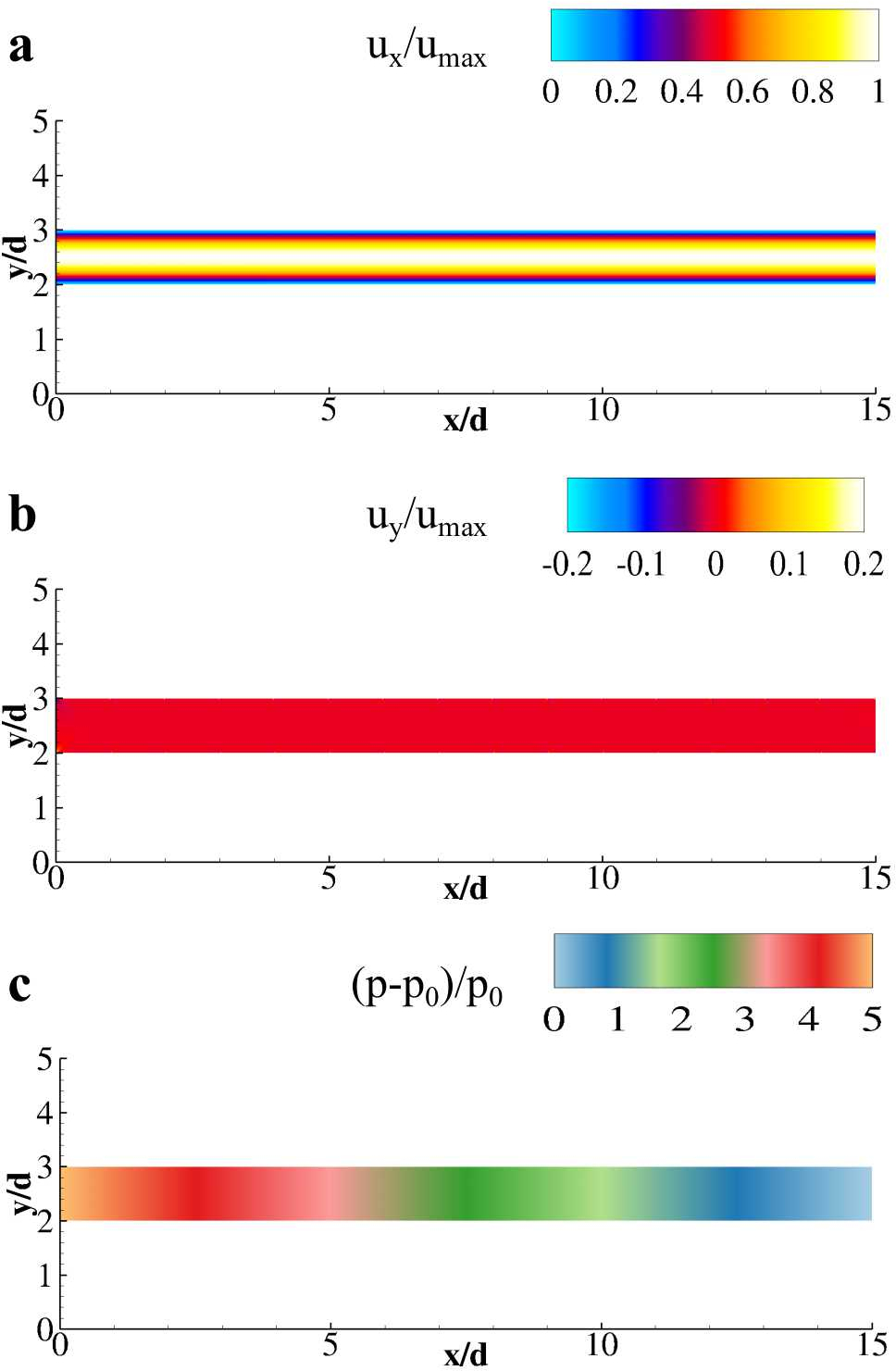}
\caption{{\bf Flow patterns in the $l/d=1$ micro-channel.} {\bf a} Contour of the longitudinal component of the velocity field. {\bf b} Contour of the vertical component of the velocity field. {\bf c} Relative pressure distribution in the computational flow field ($p_0$ is the outlet section pressure).} 
\label{SuppOnlyFluid}
\end{figure} 

\begin{figure}[h]
\centering
\includegraphics[scale=0.3]{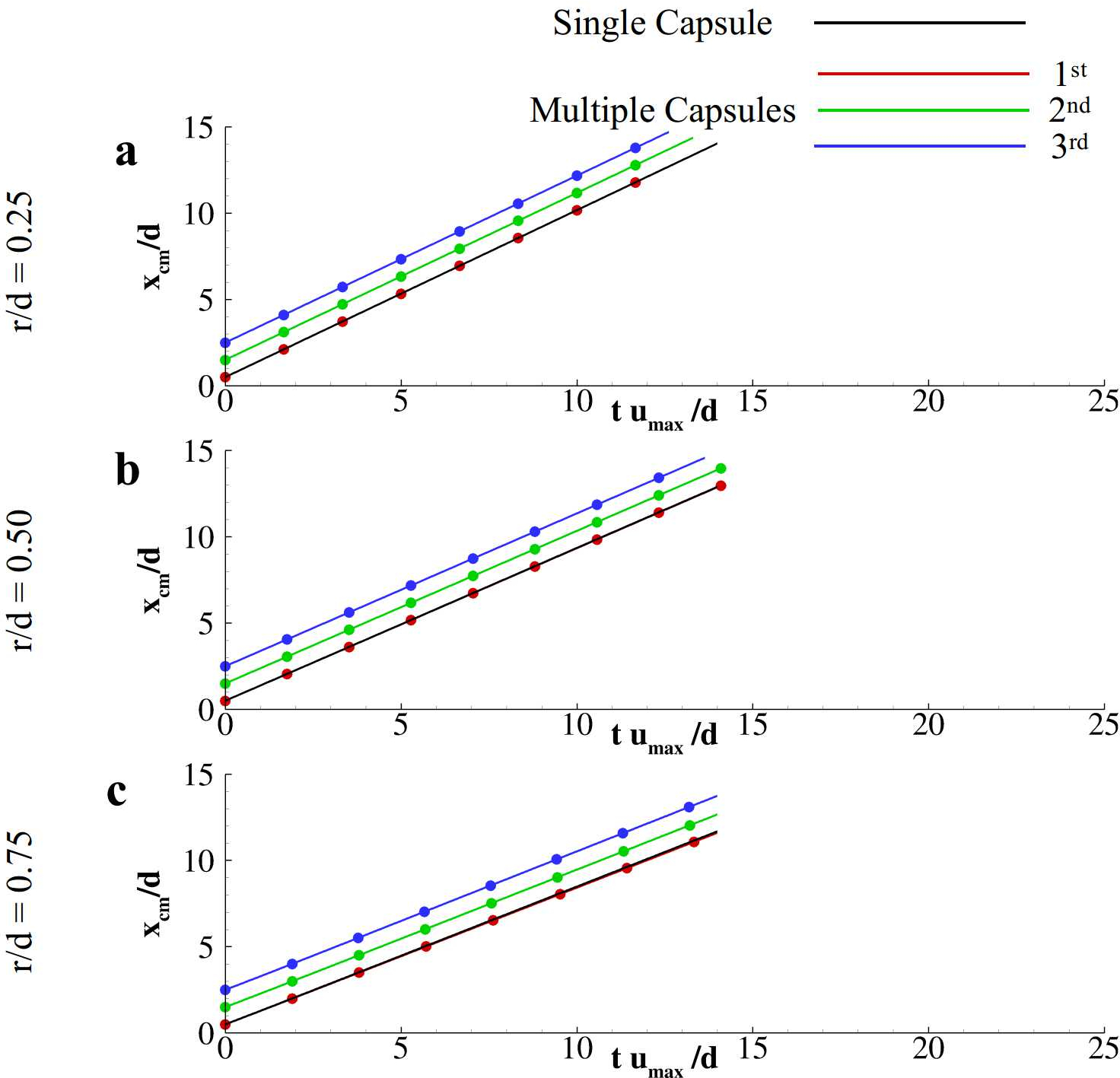}
\caption{{\bf Transport of three aligned capsules ($l/d=1.0$) in the micro-channel.} Distribution of the $x-$coordinate of the capsule centre of mass over time for $r/d=0.25$ ({\bf a}), $r/d=0.25$ ({\bf b}), and $r/d=0.25$ ({\bf c}).} 
\label{SuppMultiCaps}
\end{figure} 

\end{document}